\documentclass[aps,prd,groupedaddress,showpacs,showkeys]{revtex4}
\usepackage{epsfig}
\newcommand{\eq}{\begin{eqnarray}}
\newcommand{\en}{\end{eqnarray}}
\newcommand{\bea}{\begin{eqnarray}}
\newcommand{\eea}{\end{eqnarray}}

\newcommand{\ra}{\rangle}
\newcommand{\la}{\langle}

\begin{document}

\title{
$D_{s0}^\ast(2317)$ and $D_{s1}(2460)$ mesons in two-body $B$-meson decays
}
\author{
Amand Faessler$^1$,
Thomas Gutsche$^1$,
Sergey Kovalenko$^2$,
Valery E. Lyubovitskij$^1$
\footnote{On leave of absence
from Department of Physics, Tomsk State University,
634050 Tomsk, Russia}
\vspace*{1.2\baselineskip}}

\affiliation{$^1$ Institut f\"ur Theoretische Physik,
Universit\"at T\"ubingen,
\\ Auf der Morgenstelle 14, D-72076 T\"ubingen, Germany
\vspace*{1.2\baselineskip} \\ 
\hspace*{-1cm}$^2$
Centro de Estudios Subat\'omicos(CES),
Universidad T\'ecnica Federico Santa Mar\'\i a, \\
Casilla 110-V, Valpara\'\i so, Chile
\vspace*{0.3\baselineskip}\\}

\date{\today}

\begin{abstract}
We analyze the branching ratios of
$B \to D^{(\ast)} D_{s0}^\ast (D_{s1})$ decays using the factorization
hypothesis. The $B \to D^{(\ast)}$ transition form factors are taken
from a model-independent analysis done by Caprini, Lellouch and Neubert
based on heavy quark spin symmetry and dispersive constraints, 
including short-distance and power corrections. The leptonic decay
constants $f_{D_{s0}^\ast}$ and $f_{D_{s1}}$ are calculated assuming 
a molecular structure for the $D_{s0}^\ast$ and $D_{s1}$ mesons. 
The calculated branching ratios of $B$-meson two-body decays are 
compared with experimental data and other theoretical results.

\end{abstract}

\pacs{13.25.Hw,13.20.Fc,13.40.Hq,14.40.Lb,14.40.Nd}

\keywords{charm and bottom mesons,
leptonic and nonleptonic decays, hadronic molecule}

\maketitle

\newpage

\section{Introduction}

Presently there is strong interest to study a newly observed mesons 
and baryons in the context of a hadronic molecule interpretation 
(for overview see e.g. Ref.~\cite{Rosner:2006vc}). In the present work 
we focus on weak production properties of scalar 
$D_{s0}^\ast(2317)$ and axial $D_{s1}(2460)$ charm-strange mesons
(for a review see e.g.~\cite{Colangelo:2004vu}). 
The $D_{s0}^\ast(2317)$ meson was discovered just a few years ago by 
the BABAR Collaboration at SLAC in the inclusive $D_s^+ \pi^0$ invariant
mass distribution from $e^+ e^-$ annihilation data~\cite{Aubert:2003fg}. 
The nearby state $D_{s1}(2460)$ decaying into $D_s^\ast \pi^0$ 
was observed by the CLEO Collaboration at CESR~\cite{Besson:2003cp}. 
Both of these states have been confirmed by the Belle Collaboration 
at KEKB~\cite{Abe:2003jk}. In the interpretation of these experiments it
was suggested that the $D_{s0}^\ast(2317)$ and $D_{s1}(2460)$ mesons are
the $P$-wave charm-strange quark states with spin-parity quantum
numbers $J^P = 0^+$ and $J^P = 1^+$, respectively. In the
following the Belle~\cite{Krokovny:2003zq} and
BABAR~\cite{Aubert:2004pw} Collaborations observed the production of
$D_{s0}^\ast(2317)$, $D_{s1}(2460)$ and their subsequent strong
and radiative transitions in the nonleptonic two-body $B$ decays.  
The most recent data of BABAR and Belle on two-body 
$B$-meson decays into $D_{s0}^\ast(2317)$ and $D_{s1}(2460)$ states 
can be found in Refs.~\cite{Drutskoy:2004yv,Aubert:2006nm}. 
It is worth noting that the existing experimental information on 
the properties of $D_{s0}^\ast(2317)$ and $D_{s1}(2460)$ 
mesons~\cite{Yao:2006px} leaves quite a significant uncertainty in 
their interpretation as $J^P = 0^+$ and $J^P = 1^+$ states. 

Theoretical analysis of $B \to D^{(\ast)} D_{s0}^\ast (D_{s1})$ 
decays has been performed in different  
approaches~\cite{LeYaouanc:2001ma}-\cite{Cheng:2006dm} based on 
the factorization hypothesis, which essentially simplifies the  
calculation of the transition amplitude. The factorizable 
amplitude $B \to D^{(\ast)} D_{s0}^\ast (D_{s1})$ is given by
the product of the corresponding form factors (or their combination)
describing semileptonic $\bar B \to D^{(\ast)} \ell \bar\nu_{\ell}$
transitions and the leptonic decay constant $f_{D_{s0}^\ast}$ 
($f_{D_{s1}}$). The leptonic decay constants $f_{D_{s0}^\ast}$ and
$f_{D_{s1}}$ have been calculated directly or extracted from the
analysis of $B \to D^{(\ast)} D_{s0}^\ast (D_{s1})$ decays in
Refs.~\cite{LeYaouanc:2001ma,Cheng:2003kg,Hsieh:2003xj,%
Cheng:2003sm,Hwang:2004kg}, \cite{Cheng:2006dm}-\cite{Herdoiza:2006qv}.  
The form factors of $B \to D^{(\ast)} \ell \bar\nu_{\ell}$ transitions 
have been analyzed and calculated in various theoretical approaches   
such as: heavy quark effective theory, QCD sum rules, lattice QCD, 
different types of quark and soliton models, approaches based on 
the solution of Bethe-Salpeter and Faddeev equations, etc. 

In this paper we assume the $D_{s0}^\ast$ and $D_{s1}$ mesons to be 
hadronic molecules - bound states of $D, K$ and $D^\ast, K$ mesons,
respectively. Using this molecular picture, 
in Refs.~\cite{Faessler:2007gv,Ds1} we calculated strong and radiative 
decays of $D_{s0}^\ast$ and $D_{s1}$ 
mesons. The obtained results are in agreement with other theoretical 
approaches, e.g. the strong decay widths $D_{s0}^\ast \to D_s \pi^0$
and $D_{s1} \to D_s^\ast \pi^0$ are of the order of $10^2$ KeV and the
radiative decays $D_{s0}^\ast \to D_s^\ast \gamma$,
$D_{s1} \to D_s \gamma$, etc. are of the order of a few KeV. 
Here, using the same approach, we calculate the leptonic decay 
constants $f_{D_{s0}^\ast}$ and $f_{D_{s1}}$. For the form factors 
governing the semileptonic $\bar B \to D^{(\ast)} \ell \bar\nu_{\ell}$ 
transitions we use the model-independent results obtained 
by Caprini, Lellouch and Neubert (CLN)~\cite{Caprini:1997mu} on the 
basis of heavy quark spin symmetry and dispersive constraints, 
including short-distance and power corrections. Note, that in 
Ref.~\cite{Hwang:2004kg} the authors already used the CLN
results in their analysis of two-body $B \to D D_{s0}^\ast (D_{s1})$
transitions restricting themselves to the heavy quark limit and modes 
with a pseudoscalar $D$ meson in the final state. Using the experimental 
lower limits for the $B \to D D_{s0}^\ast (D_{s1})$ branching ratios they 
derived lower limits for the products $|a_1| f_{D_{s0}^\ast}$ and 
$|a_1| f_{D_{s1}}$, where $a_1$ is a combination of the short-distance 
Wilson coefficients~\cite{Buchalla:1995vs}-\cite{Luo:2001mc}. 

In the present paper we proceed as follows. First, in Section II, we
discuss the basic notions of our approach. We indicate and evaluate 
the effective mesonic Lagrangian for the treatment of charmed mesons
$D_{s0}^\ast(2317)$ and $D_{s1}(2460)$ as $DK$ and $D^\ast K$
bound states, respectively. Then in Section III we discuss the calculation 
of the leptonic decay constants $f_{D_{s0}^\ast}$ and $f_{D_{s1}}$.
In Section IV we present a detailed analysis of two-body bottom
meson decays $B \to D^{(\ast)} D_{s0}^\ast (D_{s1})$ applying the 
factorization hypothesis. As we already stressed before, in this analysis 
we use the model-independent CLN results~\cite{Caprini:1997mu} for 
the weak form factors defining the $B \to D^{(\ast)} \ell \bar\nu_{\ell}$ 
transitions. In Section V we give a short summary of our results.

\section{Molecular structure of $D_{s0}^{\ast \, \pm}(2317)$
and $D_{s1}^{\pm}(2460)$ mesons}

In this section we discuss the formalism for the study of the
$D_{s0}^{\ast \, \pm}(2317)$ and $D_{s1}^{\pm}(2460)$ mesons as hadronic
molecules, represented by $D K$ and $D^\ast K$  bound states, respectively. 
We adopt that the isospin, spin and parity quantum numbers of 
$D_{s0}^{\ast \, \pm}(2317)$ and $D_{s1}^{\pm}(2460)$ are: 
$I(J^P) = 0(0^+)$ and $I(J^P) = 0(1^+)$, while for their masses 
we take the values: $m_{D_{s0}^{\ast}} = 2.3173$ GeV and  
$m_{D_{s1}} = 2.4589$ GeV~\cite{Yao:2006px}. Our framework is based on 
effective interaction Lagrangians describing the couplings of 
$D_{s0}^\ast(2317)$ and $D_{s1}(2460)$ mesons to their constituents: 
\eq\label{Lagr_Ds0_Ds1}
{\cal L}_{D_{s0}^\ast}(x) &=& g_{_{D_{s0}^\ast}} \,
D_{s0}^{\ast \, -}(x) \, \int\! dy \,
\Phi_{D_{s0}^\ast}(y^2) \, D(x+w_{_{KD}} y) \,
K(x-w_{_{DK}} y) \, + \, {\rm H.c.} \,, \\
\label{Lagr_Ds0_Ds1-1}
{\cal L}_{D_{s1}}(x) &=& g_{_{D_{s1}}} \,
D_{s1}^{\mu \, -}(x) \, \int\! dy \,
\Phi_{D_{s1}}(y^2) \, D^{\ast}_\mu(x+w_{_{KD^\ast}} y) \,
K(x-w_{_{D^\ast K}} y) \, + \, {\rm H.c.} \,,
\en
where the doublets of $D^{(\ast)}$ and $K$ mesons are defined as
\eq
D =
\left(
\begin{array}{c}
D^0 \\
D^+ \\
\end{array}
\right)\,, \hspace*{1cm}
D^\ast =
\left(
\begin{array}{c}
D^{\ast \, 0} \\
D^{\ast \, +} \\
\end{array}
\right)\,, \hspace*{1cm}
K =
\left(
\begin{array}{c}
K^+ \\
K^0 \\
\end{array}
\right)\,. 
\en
The summation over isospin indices is understood. 
The molecular structure of the $D_{s0}^{\ast \pm}$
and $D_{s1}^\pm$ states is:
\eq
& &|D_{s0}^{\ast \, +}\ra \, = \,
|D^+ K^0\ra + |D^0 K^+\ra \,, \hspace*{1.15cm}
|D_{s0}^{\ast \, -}\ra \, = \,
|D^- \bar K^0\ra + |\bar D^0 K^-\ra \,, \nonumber\\
& &|D_{s1}^+\ra \, = \, |D^{\ast +} K^0\ra + |D^{\ast 0} K^+\ra \,, 
\hspace*{1cm}
|D_{s1}^-\ra \, = \, |D^{\ast -} \bar K^0\ra + |\bar D^{\ast 0} K^-\ra \,.
\en 
The correlation functions $\Phi_M$ with $M=D_{s0}^\ast$ or $D_{s1}$
characterize the finite size of the $D_{s0}^\ast(2317)$ and $D_{s1}(2460)$
mesons as $DK$ and $D^\ast K$ bound states and depend on the relative
Jacobi coordinate $y$ with, in addition, $x$ being the center of mass 
(CM) coordinate. Note, that the local limit corresponds to the substitution 
of $\Phi_M$ by the Dirac delta-function: $\Phi_M(y^2) \to \delta^4 (y)$. 
In Eqs. (\ref{Lagr_Ds0_Ds1}) and (\ref{Lagr_Ds0_Ds1-1}) we introduced 
the kinematical parameters $w_{ij}$: 
\eq
w_{ij} = \frac{m_i}{m_i + m_j}\,,
\en
where $m_D$, $m_{D^\ast}$ and $m_K$ are the masses of $D$, $D^\ast$ and
$K$ mesons. The Fourier transform of the correlation function reads
\eq
\Phi_M(y^2) \, = \,
\int\!\frac{d^4p}{(2\pi)^4}  \,
e^{-ip y} \, {\widetilde{\Phi}}_M(-p^2) \,, \hspace*{.5cm}
M = D_{s0}^\ast, D_{s1} \,.
\en 
A basic requirement for the choice of an explicit form of the correlation 
function is that it falls down sufficiently fast in the ultraviolet region 
of Euclidean space to render the Feynman diagrams ultraviolet finite. 
We adopt the Gaussian form
\eq\label{Gauss_CF}
\tilde\Phi_M(p_E^2)
\doteq \exp( - p_E^2/\Lambda^2_M)\,,
\en
for the vertex function, where $p_{E}$ is the
Euclidean Jacobi momentum. Here $\Lambda_{D_{s0}^\ast}$
is a size parameter, which parametrizes the distribution of
$D$ and $K$ mesons inside the $D_{s0}^\ast$ molecule, while 
$\Lambda_{D_{s1}}$ is the size parameter for the $D_{s1}$ molecule.
For simplicity we will use a universal scale parameter
$\Lambda_M = \Lambda_{D_{s0}^\ast} = \Lambda_{D_{s1}}$,
i.e. the same size for $D_{s0}^\ast$ and $D_{s1}$ mesons.

The coupling constants $g_{D_{s0}^\ast}$ and $g_{D_{s1}}$ are determined 
by the compositeness condition~\cite{Weinberg:1962hj,Efimov:1993ei},
which implies that the renormalization constant of the hadron
wave function is set equal to zero:
\eq\label{non-renorm-1}
Z_{D_{s0}^\ast} &=& 1 -
\Sigma^\prime_{D_{s0}^\ast}(m_{D_{s0}^\ast}^2) = 0 \,, \\
\label{non-renorm-2}
Z_{D_{s1}} &=& 1 -
\Sigma^\prime_{D_{s1}}(m_{D_{s1}}^2) = 0 \,. 
\en
Here, $\Sigma^\prime_{D_{s0}^\ast}(m_{D_{s0}^\ast}^2) =
g_{_{D_{s0}^\ast}}^2 \Pi^\prime_{D_{s0}^\ast}(m_{D_{s0}^\ast}^2)$
is the derivative of the $D_{s0}^\ast$ meson mass operator.
In the case of the $D_{s1}$ meson we have 
$\Sigma^\prime_{D_{s1}}(m_{D_{s1}}^2) 
= g_{_{D_{s1}}}^2 \Pi^\prime_{D_{s1}}(m_{D_{s1}}^2)$, which is the 
derivative of the transverse part of its mass operator
$\Sigma^{\mu\nu}_{D_{s1}}$, conventionally split into transverse
$\Sigma_{D_{s1}}$ and longitudinal $\Sigma^L_{D_{s1}}$  parts as:
\eq
\Sigma^{\mu\nu}_{D_{s1}}(p) =
g^{\mu\nu}_\perp \Sigma_{D_{s1}}(p^2) + \frac{p^\mu p^\nu}{p^2}
\Sigma^L_{D_{s1}}(p^2) \,,
\en
where
\eq
g^{\mu\nu}_\perp = g^{\mu\nu} - \frac{p^\mu p^\nu}{p^2}\,,
\hspace*{.5cm} g^{\mu\nu}_\perp p_\mu = 0\,.
\en
The mass operators of the $D_{s0}^\ast$ and $D_{s1}$ mesons
are described by the diagram of Fig.1.

Following Eqs.~(\ref{non-renorm-1}) and (\ref{non-renorm-2}) 
the coupling constants $g_{_{D_{s0}^\ast}}$ and $g_{_{D_{s1}}}$ 
can be expressed in the form:
\eq
\frac{1}{g_{_{D_{s0}^\ast}}^2} &=& \frac{2}{(4 \pi \Lambda_M)^2} \,
\int\limits_0^1 dx \int\limits_0^\infty
\frac{d\alpha \, \alpha \, P_0(\alpha, x)}{(1 + \alpha)^3}
\,\, \biggl[ - \frac{d}{dz_0} \tilde \Phi^2_M(z_0) \biggr] \,,
\label{coupling_Ds0}\\
\frac{1}{g_{_{D_{s1}}}^2} &=& \frac{2}{(4 \pi \Lambda_M)^2} \,
\int\limits_0^1 dx \int\limits_0^\infty
\frac{d\alpha \, \alpha \, P_1(\alpha, x)}{(1 + \alpha)^3}
\,\, \biggl[ \frac{1}{2 \mu_{D^\ast}^2 (1 + \alpha)}
- \frac{d}{dz_1} \biggr] \tilde \Phi^2_M(z_1)
\label{coupling_Ds1}
\en
where
\eq
P_0(\alpha, x) 
&=& \alpha^2 x(1-x) + w_{_{DK}}^2 \alpha x + w_{_{KD}}^2 \alpha (1-x)
\,, \nonumber\\
P_1(\alpha, x) 
&=& \alpha^2 x(1-x) + w_{_{D^\ast \! K}}^2 \alpha x
+ w_{_{KD^\ast}}^2 \alpha (1-x) \,, \\
z_0 &=& \mu_D^2 \alpha x +  \mu_K^2 \alpha (1-x) 
     - \frac{P_0(\alpha, x)}{1 + \alpha}  \, 
\mu_{D_{s0}^\ast}^2 \,, \nonumber\\
z_1 &=& \mu_{D^\ast}^2 \alpha x + \mu_K^2 \alpha (1-x)
   - \frac{P_1(\alpha, x)}{1 + \alpha}  \, \mu_{D_{s1}}^2 \,, 
\nonumber\\ 
\mu_M &=& \frac{m_M}{\Lambda_M} \,. \nonumber
\en 
Above expressions are valid for any functional form of the correlation 
function ${\widetilde{\Phi}}_M$.

Let us note, that the compositeness condition of the 
type~(\ref{non-renorm-1}), (\ref{non-renorm-2}) was originally applied 
to the study of the deuteron as a bound state of proton and 
neutron~\cite{Weinberg:1962hj}. Then this condition was extensively used 
in low-energy hadron phenomenology as the master equation for the 
treatment of mesons and baryons as bound states of light and heavy
constituent quarks~\cite{Efimov:1993ei}-\cite{Burdanov:1996uw}.   
In Refs.~\cite{Burdanov:2000rw,Baru:2003qq,Faessler:2007gv}  
this condition was used in the application to glueballs as bound states 
of gluons and light and heavy scalar mesons as hadronic molecules. 

\section{Leptonic decay constants $f_{_{D_{s0}^\ast}}$ and $f_{_{D_{s1}}}$}

The decay constants of the scalar $D_{s0}^\ast$ and axial $D_{s1}$ mesons 
are defined by
\eq\label{Leptonic}
& & \la 0| \bar s O_\mu  c | D_{s0}^{\ast \, +}(p) \ra
= p_\mu f_{_{D_{s0}^\ast}}
\,, \\
\label{Leptonic-1}
& & \la 0| \bar s O_\mu c | D_{s1}^+(p,\epsilon) \ra =
 \epsilon_\mu(p) \, m_{D_{s1}}  f_{_{D_{s1}}} \,,
\en
where $O_\mu = \gamma_\mu (1 - \gamma_5)$. 
Note, parity symmetry implies that only the vector component 
of the weak $V-A$ current contributes to the transition 
$D_{s0}^{\ast +} \to W^+$, while for the transition  
$D_{s1}^+ \to W^+$ only the axial component is present.  
The one-loop meson diagrams describing the matrix elements 
of Eqs.~(\ref{Leptonic}), (\ref{Leptonic-1})
are given in Figs.2a and 2b. In other words, adopting the molecular
picture for the $D_{s0}^{\ast}$ and $D_{s1}$ mesons we need to evaluate 
the two-point $DK$ loop diagram describing the transition of the 
$D_{s0}^{\ast}$  meson to the vector current and the two-point
$D^\ast K$ loop diagram corresponds to the transition of the $D_{s1}$ meson
to the axial current. The coupling of the weak $c\to s$ flavor-changing
vector current (FCVC) $V_\mu$ to the $DK$ pair and the weak $c\to s$
flavor-changing axial current (FCAC) $A_\mu$ to the $D^\ast K$ pair
can be extracted from data. In particular, the couplings are extracted 
by deriving the effective Lagrangians from the matrix elements for the 
$D \to K^{(\ast)} \ell \bar\nu_\ell$ semileptonic transitions. First, 
the matrix element for the 
$D \to K \ell \bar\nu_\ell$ transition~\cite{Wirbel:1985ji}
at zero momentum transfer $q = p - p^\prime \to 0$ is approximated by
\eq
\la K(p^\prime) | V^\mu(0) | D(p) \ra \simeq f_+^{DK}(0) (p+p^\prime)^\mu \,. 
\en
Therefore, the effective Lagrangian describing the coupling of $D$ and
$K$ mesons with the weak vector current is given by
\eq
{\cal L}_{VDK}(x) = f_+^{DK}(0) \,  V^\mu(x) \,
D(x) \, i\partial_\mu^- \, K(x) + {\rm H.c.}
\en
where $A \partial_\mu^- B = (\partial_\mu A) B - A \partial_\mu B$. 

The effective coupling of the weak axial current $A_\mu$ with $D^\ast$ 
and $K$ mesons can be related to the corresponding coupling 
with $D$ and $K^\ast$ mesons on the basis of SU(4) flavor 
symmetry arguments. Both matrix elements for the semileptonic transitions 
$D^\ast \to K \ell \bar\nu_\ell$ 
and $D \to K^\ast \ell \bar\nu_\ell$ semileptonic transitions at zero 
momentum transfer can be approximately written as:
\eq
\la K(p^\prime)| A^\mu(0) | D^\ast(p,\epsilon) \ra &\simeq&
\epsilon^{\mu} (m_{D^\ast} + m_K)  A_1^{D^\ast K}(0) \,,\\
\la K^\ast(p^\prime,\epsilon^\prime)| A^\mu(0) | D(p) \ra &\simeq&
\epsilon^{\prime \ast \mu} (m_D + m_{K^\ast})  A_1^{DK^\ast}(0) \, . 
\en
At present no experimental information on the axial form factor 
$A_1^{D^\ast K}(0)$ is available, while the form factor 
$A_1^{DK^\ast}(0)$ is known, although with significant 
uncertainties. SU(4) symmetry requires that the axial form factors 
$A_1^{D^\ast K}(0)$ and $A_1^{DK^\ast}(0)$ satisfy the relation: 
\eq\label{A1_ident}
(m_{D^\ast} + m_K)  A_1^{D^\ast K}(0) \, \equiv \,
(m_D + m_{K^\ast})  A_1^{DK^\ast}(0) \,, 
\en
which in turn allows to express the unknown form factor  
$A_1^{D^\ast K}(0)$ by $A_1^{DK^\ast}(0)$. 
The effective Lagrangian, describing the coupling of $D^\ast K$  
and $DK^\ast$ meson pairs to the weak $c\to s$ FCAC is then given as: 
\eq
{\cal L}_{AD^\ast K}(x) + {\cal L}_{ADK^\ast}(x) =
(m_D + m_{K^\ast})  \, A_1^{DK^\ast}(0) \, A^\mu(x)
\biggl\{ D^{\ast}_\mu(x)  K(x) + D(x) K^{\ast}_\mu(x) \biggr\} + {\rm H.c.}
\,.
\en
In the following we calculate the leptonic decay constants $
f_{_{D_{s0}^\ast}}$ and $f_{_{D_{s1}}}$ based on the effective 
interaction Lagrangian ${\cal L}_{\rm eff}$, which includes the couplings of
$D_{s0}^\ast$, $D_{s1}$ mesons and the weak currents (vector $V^\mu$ and axial
$A^\mu$) with $DK$ and $D^\ast K$ meson pairs:
\eq
{\cal L}_{\rm eff}(x) = {\cal L}_{D_{s0}^\ast}(x) + {\cal L}_{D_{s1}}(x)
+ {\cal L}_{VDK}(x) + {\cal L}_{AD^\ast K}(x) \,.
\en
The corresponding diagrams are given in Fig.2. After
a straightforward evaluation we obtain the following analytical 
expressions for the leptonic decay constants:
\eq 
f_{_{D_{s0}^\ast}} &=& \frac{g_{_{D_{s0}^\ast}}}{8 \pi^2} \,
f_+^{DK}(0) \, \int\limits_0^1 dx \int\limits_0^\infty
\frac{d\alpha \, \alpha}{(1 + \alpha)^2} \, \tilde \Phi_M(z_0)
\,\, \biggl[ 1 - \frac{2(w_{_{KD}} + \alpha x)}{1 + \alpha} \biggr] \,,
\label{fDs0} \\
f_{_{D_{s1}}} &=& \frac{g_{_{D_{s1}}}}{8 \pi^2} \,
\frac{m_D + m_{K^\ast}}{m_{D_{s1}}} \, A_1^{DK^\ast}(0) \,
\int\limits_0^1 dx \int\limits_0^\infty
\frac{d\alpha \, \alpha}{(1 + \alpha)^2} \, \tilde \Phi_M(z_1)
\,\, \biggl[ 1 + \frac{\alpha}{4\mu_{D^\ast}^2}  
\frac{dz_1}{d\alpha} \biggr] \label{fDs1} \,,
\en
where the coupling constants $g_{_{D_{s0}^\ast}}$ and $g_{_{D_{s1}}}$
are defined by Eqs.~(\ref{coupling_Ds0}) and (\ref{coupling_Ds1}).
We have only one model parameter in our calculations - the scale parameter
$\Lambda_M$ which was previously fixed to the value of 
$\Lambda_M = 2$ GeV~\cite{Faessler:2007gv,Ds1} from strong and 
radiative decays of $D_{s0}^\ast$ and $D_{s1}$ mesons. Further quantities, 
entering in Eqs.~(\ref{fDs0}) and (\ref{fDs1}), are chosen as follows. 
The masses of mesons we take from the PDG~\cite{Yao:2006px} 
\eq
& &m_{D_{s0}^{\ast}} = 2.3173 \ {\rm GeV}\,, \hspace*{1.7cm}
m_{D_{s1}} = 2.4589 \ {\rm GeV}\,, \nonumber\\
& &m_K \equiv m_{K^\pm} = 493.677 \ {\rm MeV}\,,       \hspace*{.5cm}
m_{K^\ast} \equiv m_{K^{\ast \, \pm}} = 891.66 \ {\rm MeV}\,, \\
& &m_D \equiv m_{D^\pm} = 1.8693 \ {\rm GeV}\,,        \hspace*{.75cm}
m_{D^\ast} \equiv m_{D^{\ast \, \pm}} = 2.010 \ {\rm GeV}\,. \nonumber
\en
For the values of the $D \to K^{(\ast)} \ell \bar\nu_\ell$ semileptonic
form factors at zero recoil, entering in our calculation, we use
the world average data~\cite{Flynn:1997ca,Ball:2006yd,Zweber:2007dy} of:
\eq\label{SLFF}
f_+^{DK}(0) = 0.75 \pm 0.05\,, \hspace*{.5cm}
A_1^{DK^\ast}(0) = 0.65 \pm 0.05 \,.
\en
Using this input we get the following results for $f_{_{D_{s0}^\ast}}$ 
and $f_{_{D_{s1}}}$:
\eq
f_{_{D_{s0}^\ast}} = 67.1 \pm 4.5  \ {\rm MeV} \,, \hspace*{.5cm}
f_{_{D_{s1}}}    =  144.5 \pm 11.1 \ {\rm MeV} \,.
\en
In Table 1 we summarize the present results for $f_{_{D_{s0}^\ast}}$ and
$f_{_{D_{s1}}}$ obtained in different approaches (either on the basis 
of hadronic models or from the analysis of experimental data on two-body 
$B$-meson decays). Our results are in agreement with the predictions 
of Refs.~\cite{Hwang:2004kg,Cheng:2003sm,Cheng:2006dm,Cheng:2003kg},
especially with the lower limits derived from an analysis of the
branching ratios of $B \to D^{(\ast)} D_{s0}^\ast (D_{s1})$
decays~\cite{Hwang:2004kg,Cheng:2006dm}.

At this point it is worth discussing the heavy quark limit (HQL) for the 
leptonic decay constants $f_{_{D_{s0}^\ast}}$ and $f_{_{D_{s1}}}$, 
where the masses of $D$, $D^\ast$, $D_{s0}^\ast$ and $D_{s1}$ mesons together
with the charm quark mass $m_c$ approach infinity. In the HQL the $D^{(\ast)}$ 
mesons in the $D_{s0}^\ast (D_{s1})$ hadronic molecules 
move to the center of mass and are surrounded by a light $K$ meson
in analogy with the heavy-light $Q \bar q$ mesons.
It is known (see e.g. discussion in Refs.~\cite{Cheng:2003sm,Hwang:2004kg})
that in the two-quark picture for the $D_{s0}^\ast$, $D_{s1}$ mesons 
HQL leads to degenerate values for these couplings
\eq\label{scale-1}
f_{_{D_{s0}^\ast}} \equiv f_{_{D_{s1}}} \sim \frac{1}{\sqrt{m_c}} \,.
\en
In the present molecular approach we can also guarantee that the couplings are 
degenerate in the HQL, but the scaling law is different from the $1/\sqrt{m_c}$ 
behavior. In a first step we apply the HQL to the coupling constants 
$g_{_{D_{s0}^\ast}}$ and $g_{_{D_{s1}}}$, which are degenerate:
\eq\label{HQL_nonl}
& &\frac{1}{g_{_{D_{s0}^\ast}}^2} \equiv \frac{1}{g_{_{D_{s1}}}^2}
= \frac{I_0}{(4 \pi m_c)^2}  \,, \\
& &I_0 = \int\limits_0^\infty \frac{d\alpha}{1 + \alpha} \,\,
\Phi^2_M(\mu_K^2 \alpha) \,.  \nonumber
\en
The structure integrals
$\int\limits_0^1 dx \int\limits_0^\infty d\alpha \cdots$
entering in Eqs.~(\ref{fDs0}) and (\ref{fDs1}) are also degenerate 
and equal 
\eq
I_1 =  \frac{\Lambda_M}{m_c} \, \int\limits_0^\infty
\frac{d\alpha \, \sqrt{\alpha}}{1 + \alpha} \,\,
\Phi^2_M(\mu_K^2 \alpha) \,.
\en
Finally, the leptonic decay constants $f_{_{D_{s0}^\ast}}$ and
$f_{_{D_{s1}}}$ in the HQL are given by:
\eq\label{LepC_HQL}
& & f_{_{D_{s0}^\ast}} \, = \, \frac{\Lambda_M}{2 \pi}
\, g_{_{VDK}}^{hql} \, \frac{I_1}{\sqrt{I_0}} \,, \\
\label{LepC_HQL-1}
& & f_{_{D_{s1}}} \, = \, \frac{\Lambda_M}{2 \pi} \,
g_{_{AD^\ast K}}^{hql} \, \frac{I_1}{\sqrt{I_0}} \,, 
\en
where we introduced the effective $VDK$ and $AD^\ast K$ couplings
in the HQL: $g_{_{VDK}}^{hql}$ and $g_{_{AD^\ast K}}^{hql}$. 
From Eqs.~(\ref{LepC_HQL}) and (\ref{LepC_HQL-1}) we deduce that in 
the HQL $f_{_{D_{s0}^\ast}}$ and $f_{_{D_{s1}}}$ in the HQL limit do 
not depend on $m_c$ at all, unlike as in Eq.~(\ref{scale-1}),  
and are degenerate for $g_{_{VDK}}^{hql} = g_{_{AD^\ast K}}^{hql}$. 
The latter condition can be eventually fulfilled, e.g. at finite 
masses (see Eqs.~(\ref{fDs0}) and (\ref{fDs1})) the ratio
$g_{_{AD^\ast K}}/g_{_{VDK}}$ is close to 1:
\eq
\frac{g_{_{AD^\ast K}}}{g_{_{VDK}}} \, = \,
\frac{m_D + m_{K^\ast}}{m_{D_{s1}}} \,
\frac{A_1^{DK^\ast}(0)}{f_+^{DK}(0)}
\simeq 1 \,.
\en 
To give an estimate for the absolute values of the leptonic decay constants
$f_{_{D_{s0}^\ast}}$ and $f_{_{D_{s1}}}$ in the HQL we vary
the coupling constants $g_{_{VDK}}^{hql} = g_{_{AD^\ast K}}^{hql}$
in the region $0.75 \pm 0.25$. The corresponding result is:
\eq
f_{_{D_{s0}^\ast}} = f_{_{D_{s1}}} = 205.2 \pm 68.4 \ {\rm MeV} \,.
\en 
The following comment is in order. As was already stressed 
in Ref.~\cite{Hwang:2004kg} the HQL does not give 
a reasonable approximation for the $P$-wave $D_{sJ}$ meson system. 
According to the data on the two-body decays 
$B \to D^{(\ast)} D_{s0}^\ast (D_{s1})$ the physical value of 
the decay constant $f_{_{D_{s1}}}$ should be about twice as large 
as $f_{_{D_{s0}^\ast}}$~\cite{Hwang:2004kg,Cheng:2006dm}.

\section{Weak two-body decays $B \to D^{(\ast)} D_{s0}^\ast (D_{s1})$}

In this section we give the predictions for the branching ratios of
$B \to D^{(\ast)} D_{s0}^\ast (D_{s1})$ decays. For this purpose we use 
leptonic decay constants $f_{_{D_{s0}^\ast}}$, $f_{_{D_{s1}}}$ and, 
in addition, model-independent results for the form factors of
$B \to D^{(\ast)} \ell \bar\nu_{\ell}$ transitions obtained 
by Caprini, Lellouch and Neubert (CLN)~\cite{Caprini:1997mu}. 
Latter derivations are based on heavy quark spin symmetry,
dispersive constraints, including short-distance and power corrections.
Note, that in Ref.~\cite{Hwang:2004kg} the authors already used the CLN
results in their analysis of two-body $B \to D D_{s0}^\ast (D_{s1})$
transitions, restricting to modes containing the pseudoscalar
meson $D$ in the final state.

Working with the factorization approximation we first write down the
factorizable amplitudes for the four decay modes
$\bar B^0 \to D^+ D_{s0}^{\ast -}$, $D^+ D_{s1}^-$,
$D^{\ast +} D_{s0}^{\ast -}$ and $D^{\ast +} D_{s1}^-$ as expressed 
in terms of matrix elements of the semileptonic 
$B \to D^{(\ast)} \ell \bar\nu_{\ell}$ 
and the leptonic $D_{s0}^{\ast}$, $D_{s1}$ transitions and 
(see details in Refs.~\cite{Cheng:2003sm,Hwang:2004kg,Cheng:2006dm}):
\eq\label{Fact_amplitude}
M(\bar B^0 \to D^{+} D_{s0}^{\ast -}) &=& G_{\rm eff}
\la D_{s0}^{\ast \, -}(q) | \bar s O_\mu c | 0 \ra \,
\la D^{+}(p^\prime) | \bar c O^\mu b | \bar B^0(p) \ra \,,\\
M(\bar B^0 \to D^{\ast +} D_{s0}^{\ast -}) &=& G_{\rm eff}
\la D_{s0}^{\ast \, -}(q) | \bar s O_\mu c | 0 \ra \,
\la D^{\ast +}(p^\prime,\epsilon^\prime) 
| \bar c O^\mu b | \bar B^0(p) \ra \,,\\
M(\bar B^0 \to D^{+} D_{s1}^{-}) &=& G_{\rm eff}
\la D_{s1}^{-}(q,\epsilon) | \bar s O_\mu c | 0 \ra \,
\la D^{+}(p)
| \bar c O^\mu b | \bar B^0(p) \ra \,,\\
M(\bar B^0 \to D^{\ast +} D_{s1}^{-}) &=& G_{\rm eff}
\la D_{s1}^{-}(q,\epsilon) | \bar s O_\mu c | 0 \ra \,
\la D^{\ast +}(p^\prime,\epsilon^\prime)
| \bar c O^\mu b | \bar B^0(p) \ra \,.
\en
Here 
\eq
G_{\rm eff} = \frac{G_F}{\sqrt{2}} V_{cb} V_{cs}^\ast a_1 \,
\en 
and $a_1 = c_2 + c_1/N_c$ is the 
combination of the short-distance Wilson
coefficients $c_1$ and $c_2$~\cite{Buchalla:1995vs}-\cite{Luo:2001mc}. 
Using the Wirbel-Stech-Bauer (WSB) 
decomposition~\cite{Wirbel:1985ji,Neubert:1993mb} of the hadronic 
$B \to D^{(\ast)}$ matrix elements
\eq
\la D^+(p^\prime) | V^\mu | \bar B^0(p) \ra &=&
\biggl\{ (p+p^\prime)^\mu - \frac{m_B^2 - m_D^2}{q^2} q^\mu \biggr\} F_1(q^2)
+ \frac{m_B^2 - m_D^2}{q^2} q^\mu F_0(q^2) \,, \\
\la D^{\ast +}(p^\prime,\epsilon^\prime) | V^\mu | \bar B^0(p) \ra &=&
\frac{2i\epsilon^{\mu\nu\alpha\beta}}{m_B+m_{D^\ast}}
\, \epsilon^{\prime \ast}_\nu \, p^\prime_\alpha \, p_\beta \, V(q^2)\,, \\
\la D^{\ast +}(p^\prime,\epsilon^\prime)| A^\mu | \bar B^0(p) \ra &=&
\epsilon^{\prime \ast \mu} (m_B + m_{D^\ast})  A_1(q^2) \nonumber\\
&-& \frac{\epsilon^{\prime \ast } q}{m_B + m_{D^\ast}}
(p+p^\prime)^\mu A_2(q^2)
+ 2 m_{D^\ast} q^\mu \frac{\epsilon^{\prime \ast } q}{q^2}
\biggl\{ A_0(q^2) - A_3(q^2) \biggr\} \,, 
\en
where $V^\mu = \bar c \gamma^\mu b$,
$A^\mu = \bar c \gamma^\mu \gamma^5 b$ and
\eq
A_3(q^2) = \frac{m_B+m_{D^\ast}}{2m_{D^\ast}} A_1(q^2)
         - \frac{m_B-m_{D^\ast}}{2m_{D^\ast}} A_2(q^2)
\en
we arrive at~\cite{Cheng:2003sm,Hwang:2004kg,Cheng:2006dm}:
\eq
M(\bar B^0 \to D^+ D_{s0}^{\ast -}) &=& G_{\rm eff}  \,
f_{_{D_{s0}^\ast}} \, (m_B^2 - m_D^2) \, F_0(m_{D_{s0}^\ast}^2)\,, \\
M(\bar B^0 \to D^+ D_{s1}^-) &=& G_{\rm eff} \,
f_{_{D_{s1}}} \, m_{D_{s1}} \, \epsilon^\ast ( p + p^\prime ) \,
F_1(m_{D_{s1}}^2) \,,\\
M(\bar B^0 \to D^{\ast +} D_{s0}^{\ast -}) &=& 2 \, G_{\rm eff} \,
f_{_{D_{s0}^\ast}} \, m_{D^\ast} \, \epsilon^{\prime \ast}p \,
A_0(m_{D_{s0}^\ast}^2) \,,\\
M(\bar B^0 \to D^{\ast +} D_{s1}^-) &=& G_{\rm eff} \,
f_{_{D_{s1}}} \, m_{D_{s1}} \, \biggl\{
\frac{2 \, i \epsilon^{\mu\nu\alpha\beta}}{m_B+m_{D^\ast}} \,
\epsilon^\ast_\mu \epsilon^{\prime \ast}_\nu \,
p^\prime_\alpha \, p_\beta \, V(m_{D_{s1}}^2) \nonumber\\
&+& \epsilon^{\prime \ast} \epsilon^{\ast} (m_B + m_{D^\ast})
A_1(m_{D_{s1}}^2) - \frac{2 \, \epsilon^{\prime \ast} p  \,
\epsilon^\ast p}{m_B + m_{D^\ast}}
A_2(m_{D_{s1}}^2)  \biggr\} \,. 
\en
In the following it is convenient to express the WSB set of form factors
through a set of form factors $h_i(w)$ depending on the kinematical
variable $w = v \cdot v^\prime$, the scalar product of the four-velocities
of $\bar B$ and $D^{(\ast)}$
mesons~\cite{Falk:1992wt,Neubert:1993mb,Caprini:1997mu}:
\eq
\la D(v^\prime) | V^\mu | \bar B(v) \ra &=&
h_+(w) (v+v^\prime)^\mu + h_-(w) (v-v^\prime)^\mu \,, \\
\la D^\ast(v^\prime,\epsilon^\prime) | V^\mu | \bar B(v) \ra &=&
i h_V(w) \epsilon^{\mu\nu\alpha\beta}
\, \epsilon^{\prime \ast}_\nu \, v^\prime_\alpha \, v_\beta \,, \\
\la D^\ast(v^\prime,\epsilon^\prime)| A^\mu | \bar B(v) \ra &=&
h_{A_1}(w) \, (w+1) \epsilon^{\prime \ast \mu}
- \biggl[ h_{A_2}(w) v^\mu + h_{A_3}(w) v^{\prime\mu} \biggr]
\epsilon^{\prime \ast} v \,,
\en
where the meson states $| M(v) \ra $ obey the mass-independent
normalization condition
\eq
\la M(v^\prime) | M(v) \ra =
2 v^0 (2\pi)^3 \delta^3({\bf p} - {\bf p^\prime})\,, 
\en
instead of the relativistic one used before:
\eq
\la M(p^\prime) | M(p) \ra =
2 p^0 (2\pi)^3 \delta^3({\bf p} - {\bf p^\prime}) \,. 
\en
The WSB form factors are expressed in terms of the 
$h_i(w)$ form factors as:
\eq
& &F_1(q^2) = \frac{1+r}{2\sqrt{r}} \, V_1(w) \,, \\
& &F_0(q^2) = \frac{2\sqrt{r}}{1+r} \, \frac{w+1}{2} \,  
V_1(w) \, R(w) \,, \\
& &V(q^2)   = \frac{1+r^\ast}{2\sqrt{r^\ast}}
\, h_{A_1}(w)    \, R_1(w) \,, \\
& &A_1(q^2) = \frac{2\sqrt{r^\ast}}{1+r^\ast}
\, \frac{w+1}{2} \, h_{A_1}(w) \,, \\
& &A_2(q^2) = \frac{1+r^\ast}{2\sqrt{r^\ast}} \,  h_{A_1}(w) \, R_2(w) \,, \\
& &A_0(q^2) = \frac{1}{2 \sqrt{r^\ast}} \, h_{A_1}(w) \, R_0(w) \,,
\en
where $r = m_D/m_B$, $r^\ast = m_{D^\ast}/m_B$ and
\eq
V_1(w) = G(w) = h_+(w) - \frac{1-r}{1+r} h_-(w) \,. 
\en
Here we use the well-known form factor ratios $R_1(w)$ and
$R_2(w)$~\cite{Neubert:1993mb,Caprini:1997mu}:
\eq
R_1(w) = \frac{h_V(w)}{h_{A_1}(w)}\,, \hspace*{2.5cm}
R_2(w) = \frac{h_{A_3}(w) + r^\ast h_{A_2}(w)}{h_{A_1}(w)}\,.
\en
In the literature the form factor $V_1(w)$ is also denoted
as $G(w)$. In order to express all $h_i(w)$ form factors,  
completely defining the $B \to D^{(\ast)}$ transitions, 
through the two form factors $V_1(w)$ and $h_{A_1}(w)$
we introduce the additional ratios $R(w)$, $R_2^\ast(w)$:
\eq
R(w) = \frac{\displaystyle{h_+(w)-\frac{1+r}{1-r}\frac{w-1}{w+1} h_-(w)}}  
{\displaystyle{h_+(w)-\frac{1-r}{1+r} h_-(w)}}\,, \hspace*{2cm}
R_2^\ast(w) = \frac{h_{A_3}(w) - r^\ast h_{A_2}(w)}{h_{A_1}(w)}\,.
\en
and $R_0(w)$, which is just the combination of $R_2(w)$ and $R^\ast_2(w)$: 
\eq
R_0(w) = (w+1) [ 1 - R_2^\ast(w) ]
       + \frac{(1+r^\ast)^2}{2r^\ast} \biggl[ R_2^\ast(w)
       - \frac{1-r^\ast}{1+r^\ast} R_2(w) \biggr] \,.
\en
For the functions $V_1(w) = G(w)$, $h_{A_1}(w)$, $R_0(w)$, $R_1(w)$, $R_2(w)$
and $R_2^\ast(w)$ we use the model-independent results derived
by Caprini, Lellouch and Neubert (CLN)~\cite{Caprini:1997mu}: 
\eq
& &\frac{G(w)}{G(1)} = 1 - 8 \rho^2_G z + (51 \rho^2_G - 10) z^2
- (252 \rho^2_G - 84) z^3 \,, \\
& &\frac{h_{A_1}(w)}{h_{A_1}(1)} =
1 - 8 \rho^2_{h_{A_1}} z + (53  \rho^2_{h_{A_1}} - 15) z^2
- (231  \rho^2_{h_{A_1}} - 91) z^3 \,, \\
& &R(w)        = 1.004 - 0.007 (w-1) + 0.002 (w-1)^2 \,, \\
& &R_1(w)      = 1.27 - 0.12 (w-1) + 0.05 (w-1)^2\,, \\
& &R_2(w)      = 0.80 + 0.11 (w-1) - 0.06 (w-1)^2\,, \\
& &R_2^\ast(w) = 1.15 - 0.07 (w-1) - 0.11 (w-1)^2\,,
\en
where
\eq
z = \frac{\sqrt{w+1} - \sqrt{2}}{\sqrt{w+1} + \sqrt{2}} \,. 
\en
In the derivation of the expressions for the ratios $R_0(w)$ and $R_2^\ast(w)$
we used the results of Ref.~\cite{Caprini:1997mu} (see Appendix). Note,
that $R(w)$ is very close to 1 for $w$ in the interval
$1 \le w \le w_{\rm max} = (m_B^2 + m_D^2)/(2m_Bm_D)$.

The two-body decay widths $\Gamma(\bar B \to D^{(\ast )} D_{s0}^{\ast -})$
and $\Gamma(\bar B \to D^{(\ast)} D_{s1}^{-})$, given in terms of the CLN 
form factors, are expressed by the formulas 
(see e.g. Ref.~\cite{Hwang:2004kg}):
\eq
& &\Gamma(\bar B \to D D_{s0}^{\ast -}) =
\frac{G_{\rm eff}^2}{8 \pi m_B} \, f_{_{D_{s0}^\ast}}^2 \,
(m_B - m_D)^2 \, m_D^2 \, (w_1+1)^2 \, \sqrt{w_1^2 - 1}
\,\, [G(w_1) \, R(w_1)]^2  \, \\
& &\Gamma(\bar B \to D D_{s1}^-) =
\frac{G_{\rm eff}^2}{8 \pi m_B} \, f_{_{D_{s1}}}^2 \,
(m_B + m_D)^2 \, m_D^2 \, (w_2^2 - 1)^{3/2} \,\, G^2(w_2)
\, \\
& &\Gamma(\bar B \to D^{\ast} D_{s0}^{\ast -}) =
\frac{G_{\rm eff}^2}{8 \pi m_B} \, f_{_{D_{s0}^\ast}}^2 \,
m_B^2 \, m_{D^\ast}^2 \, (w_3^2 - 1)^{3/2}
\,\, [h_{A_1}(w_3) \, R_0(w_3)]^2  \, \\
& &\Gamma(\bar B \to D^{\ast} D_{s1}^-)  =
\frac{G_{\rm eff}^2}{8 \pi m_B} \, f_{_{D_{s1}}}^2 \,
(m_B - m_{D^\ast})^2 \, m_{D^\ast}^2 \, (w_4 + 1)^2 \, \sqrt{w_4^2 - 1}
\,\, h_{A_1}^2(w_4) \, \beta_{h_{A_1}}(w_4)\,,
\en
where the kinematical variables $w_i$ are defined as follows:
\eq
& &w_1 = \frac{m_B^2 + m_D^2 - m_{D_{s0}^\ast}^2}{2 m_B m_D}\,, \hspace*{.6cm}
   w_2 = \frac{m_B^2 + m_D^2 - m_{D_{s1}}^2}{2 m_B m_D}\,, \nonumber\\
& &w_3 = \frac{m_B^2 + m_{D^\ast}^2
- m_{D_{s0}^\ast}^2}{2 m_B m_{D^\ast}}\,,
\hspace*{.5cm}
w_4 = \frac{m_B^2 + m_{D^\ast}^2 - m_{D_{s1}}^2}{2 m_B m_{D^\ast}}\,,
\en
and
\eq
\beta_{h_{A_1}}(w) = 2 \frac{1 - 2 w r^\ast + r^{\ast 2}}{(1 - r^\ast)^2}
\biggl[ 1 + \frac{w-1}{w+1} R_1^2(w)\biggr] +
\biggl[ 1 + \frac{w-1}{1-r^\ast} (1 - R_2(w))\biggr]^2 \,. 
\en
Note that the product $h_{A_1}^2(w) \, \beta_{h_{A_1}}(w)$ defines
the well-known function $F(w)$~\cite{Neubert:1993mb,Caprini:1997mu}, 
which governs the semileptonic transition $\bar B \to D^\ast \ell \bar\nu_\ell$
(see e.g. Eq.(35) in Ref.~\cite{Caprini:1997mu}):
\eq
h_{A_1}^2(w) \, \beta_{h_{A_1}}(w) \equiv F^2(w) \,
\biggl[ 1 + \frac{4w}{w+1} \, \frac{1 -
2 w r^\ast + r^{\ast 2}}{(1-r^\ast)^2} \biggr] \,.
\en
In the numerical calculation we use the value for the 
Cabibbo-Kobayashi-Maskawa (CKM) matrix 
element $V_{cs} = 0.97296 \pm 0.00024$ from the 2006 global 
fit~\cite{Yao:2006px,Zweber:2007dy} and the 2006 averaged values
from the Heavy Flavor Averaging Group (HFAG)~\cite{HFAG:2006}:
\eq
& &|V_{cb}| \, F(1) = (36.2 \pm 0.8) \times 10^{-3}\,, \hspace*{.5cm}
\rho_{F}^2    = 1.19  \pm  0.06 \,, \\
& & |V_{cb}| \, G(1) = (42.4 \pm 4.4) \times 10^{-3}\,, \hspace*{.5cm}
\rho_{G}^2    = 1.17  \pm 0.18 \,,
\en
where the normalization of $F(w)$ and its slope $\rho_F^2$ 
at $w=1$ are related to the characteristics of the $h_{A_1}(w)$ 
form factor as~\cite{Neubert:1993mb,Caprini:1997mu}:
\eq
F(1) \equiv h_{A_1}(1)\,, \hspace*{1cm}
\rho_F^2 \simeq \rho_{h_{A1}}^2 - 0.21 \,.
\en
For the parameter $a_1$ we use the value of 1.05 from 
Ref.~\cite{Luo:2001mc}. A detailed discussion concerning the choice 
of $a_1$ can be found in Ref.~\cite{Cheng:1998kd}.

In Table 2 we present our predictions for the branching ratios 
of two-body decays $B \to D^{(\ast)} D_{s0}^\ast (D_{s1})$. 
For the data we use the averaged lower limits from 
PDG~\cite{Yao:2006px} and in addition for the modes with 
$D_{s1}(2460)$ meson in the final state the direct results of 
the BABAR Collaboration~\cite{Aubert:2006nm}. 
Our predictions are in good agreement with the experimental 
data except for the marginal situation in the case of 
$\bar B^0 \to D_{s0}^{\ast -} D^{\ast +}$ decay, 
where our prediction is slightly lower than the experimental 
limit. In Table 3 we present the results for the ratios 
$\Gamma(B \to D^\ast D_{s0}^\ast)/\Gamma(B \to D D_{s0}^\ast)$ 
and $\Gamma(B \to D^\ast D_{s1})/\Gamma(B \to D D_{s1})$. 
We also use the compilation of experimental data and theoretical
results within the covariant light-front (CLF) approach~\cite{Cheng:2003sm}
summarized in Table 10 of Ref.~\cite{Cheng:2006dm}. Here, our predictions are 
in good agreement with the existing experimental data. In comparison with 
the CLF approach our predictions are lower, although not significantly.

\section{Conclusion}

We considered the new charmed-strange mesons
$D_{s0}^\ast(2317)$ and $D_{s1}(2460)$ as hadronic molecules, that is 
$DK$ and $D^\ast K$ bound states, respectively. Using an effective 
Lagrangian approach describing the coupling of
$D_{s0}^\ast(2317)$ and $D_{s1}(2460)$ states with their constituents
we determined the leptonic decay constants $f_{D_{s0}^\ast}$ and
$f_{D_{s1}}$. Then we presented a detailed analysis of two-body bottom
mesons decays $B \to D^{(\ast)} D_{s0}^\ast (D_{s1})$ using the 
factorization hypothesis and model-independent results for the form
factors of the $B \to D^{(\ast)} \ell \bar\nu_{\ell}$ transitions 
obtained by Caprini, Lellouch and Neubert (CLN)~\cite{Caprini:1997mu}.  
The decay widths are derived in terms of the CLN form factors.
Calculated branching ratios for $B \to D^{(\ast)} D_{s0}^\ast (D_{s1})$
decays are in good agreement with experimental results 
(or the lower limits), supporting a possible interpretation of 
$D_{s0}^\ast(2317)$ and $D_{s1}(2460)$ as hadronic molecules. 

\begin{acknowledgments}

This work was supported by the DFG under contracts FA67/31-1 and
GRK683 and by CONICYT (Chile) under grants
FONDECYT No.1030244 and PBCT/No.285/2006.
This research is also part of the EU Integrated
Infrastructure Initiative Hadronphysics project under contract
number RII3-CT-2004-506078 and President grant of Russia
"Scientific Schools"  No. 5103.2006.2.

\end{acknowledgments}

\begin{center}
{\bf Table 1.}
Leptonic decay constants $f_{_{D_{s0}^\ast}}$
and $f_{_{D_{s1}}}$.

\vspace*{.25cm}

\def\arraystretch{1.2}
\begin{tabular}{|l|l|l|}
\hline
\hspace*{.5cm} Approach \hspace*{.5cm}
& \hspace*{.5cm} $f_{_{D_{s0}^\ast}}$ (MeV) \hspace*{.5cm}
& \hspace*{.5cm} $f_{_{D_{s1}}}$ (MeV) \hspace*{.5cm} \\
\hline
\,\,\,\,\, Ref.~\cite{Colangelo:2005hv}
& \,\,\,\,\,\,\,\, 225 $\pm$ 25 \,\,\,\,\,
& \,\,\,\,\,\,\,\, 225 $\pm$ 25 \,\,\,\,\, \\
\hline
\,\,\,\,\, Ref.~\cite{Jugeau:2005yr}
& \,\,\,\,\,\,\,\, 206 $\pm$ 120 \,\,\,\,\,
& \\
\hline
\,\,\,\,\, Ref.~\cite{Herdoiza:2006qv}
& \,\,\,\,\,\,\,\, 200$\pm$50 \,\,\,\,\,
& \\
\hline
\,\,\,\,\, Ref.~\cite{Colangelo:1991ug}
& \,\,\,\,\,\,\,\, 170 $\pm$ 20 \,\,\,\,\,
& \,\,\,\,\,\,\,\, 247 $\pm$ 37 \,\,\,\,\, \\
\hline
\,\,\,\,\, Ref.~\cite{Hsieh:2003xj}
& \,\,\,\,\,\,\,\, 138$\pm$16\,\,\,\,\,
& \,\,\,\,\,\,\,\, 259$\pm$13 \,\,\,\,\,  \\
\hline
\,\,\,\,\, Ref.~\cite{Veseli:1996yg}
& \,\,\,\,\,\,\,\, 110$\pm$18\,\,\,\,\,
& \,\,\,\,\,\,\,\, 233$\pm$31 \,\,\,\,\,  \\
\hline
\,\,\,\,\, Ref.~\cite{Hwang:2004kg}
& \,\,\,\,\,\,\,\, $>$ (74 $\pm$ 11)/$|a_1|$ \,\,\,\,\,
& \,\,\,\,\,\,\,\, $>$ (166 $\pm$ 20)/$|a_1|$ \,\,\,\,\,  \\
\hline
\,\,\,\,\, Ref.~\cite{Cheng:2003sm}
& \,\,\,\,\,\,\,\, 71 \,\,\,\,\,
& \,\,\,\,\,\,\,\, 117 \,\,\,\,\,  \\
\hline
\,\,\,\,\, Ref.~\cite{Cheng:2003sm}
& \,\,\,\,\,\,\,\, $60 \pm 13$  \,\,\,\,\,
& \,\,\,\,\,\,\,\, $150 \pm 40$ \,\,\,\,\,  \\
\hline
\,\,\,\,\, Ref.~\cite{Cheng:2006dm}
& \,\,\,\,\,\,\,\, $>$ (58 - 86)/$|a_1|$ \,\,\,\,\,
& \,\,\,\,\,\,\,\, $>$ (90 - 228)/$|a_1|$ \,\,\,\,\,  \\
\hline
\,\,\,\,\, Ref.~\cite{Cheng:2003kg}
& \,\,\,\,\,\,\,\, 67$\pm$13 \,\,\,\,\,
& \\
\hline
\,\,\,\,\, Ref.~\cite{LeYaouanc:2001ma}
& \,\,\,\,\,\,\,\, 44 \,\,\,\,\,
& \,\,\,\,\,\,\,\, 41 \,\,\,\,\, \\
\hline
\,\,\,\,\, Our results \,\,\,\,\,
& \,\,\,\,\,\,\,\,  67.1 $\pm$ 4.5  \,\,\,\,\,
& \,\,\,\,\,\,\,\, 144.5 $\pm$ 11.1 \,\,\,\,\, \\
\hline
\end{tabular}
\end{center}

\vspace*{.5cm}

\begin{center}
{\bf Table 2.}
Branching ratios of $B \to D^{(\ast)} D_{s0}^\ast (D_{s1})$ decays
(in units of 10$^{-3}$).

\vspace*{.25cm}

\def\arraystretch{1.2}
\begin{tabular}{|l|l|l|l|}
\hline
\hspace*{.5cm} Mode \hspace*{.5cm}
& Data (averaged)~\cite{Yao:2006px} & BABAR~\cite{Aubert:2006nm}
& Our results \\
\hline
$B^- \to D_{s0}^{\ast -} D^0$             & $>$ $0.74^{+0.23}_{-0.19}$
                                          & & $1.03 \pm 0.14$\\
\hline
$\bar B^0 \to D_{s0}^{\ast -} D^+$        & $>$ $0.97^{+0.41}_{-0.34}$
                                          & & $0.96 \pm 0.13$\\
\hline
$B^- \to D_{s1}^- D^0$                    & $>$ $1.4^{+0.6}_{-0.5}$
                                          & $4.3 \pm 1.6 \pm 1.3$
                                          & $2.54 \pm 0.39$\\
\hline
$\bar B^0 \to D_{s1}^- D^+$               & $>$ $2.0^{+0.6}_{-0.5}$
                                          & $2.6 \pm 1.5 \pm 0.74$
                                          & $2.36 \pm 0.36$\\
\hline
$B^- \to D_{s0}^{\ast -} D^{\ast 0}$      & $>$ $0.9 \pm 0.6^{+0.4}_{-0.3}$
                                          & & $0.50 \pm 0.07$\\
\hline
$\bar B^0 \to D_{s0}^{\ast -} D^{\ast +}$ & $>$ $1.5 \pm 0.4^{+0.5}_{-0.4}$
                                          & & $0.47 \pm 0.06$\\
\hline
$B^- \to D_{s1}^- D^{\ast 0}$             & $>$ $5.5 \pm 1.2^{+2.2}_{-1.6}$7.6
                                          & $11.2 \pm 2.6 \pm 2.0$
                                          & $7.33 \pm 1.12$ \\
\hline
$\bar B^0 \to D_{s1}^- D^{\ast +}$        & $>$ $7.6 \pm 1.7^{+3.2}_{-2.4}$
                                          & $8.8 \pm 2.0 \pm 1.4$
                                          & $6.85 \pm 1.05$ \\
\hline
\hline
\end{tabular}
\end{center}

\vspace*{.5cm}

\begin{center}
{\bf Table 3.} Ratios
$M D^\ast/M D = \Gamma(B \to M D^\ast)/\Gamma(B \to M D)$
for $M = D_{s0}^\ast, D_{s1}$.

\vspace*{.25cm}

\def\arraystretch{1.2}
\begin{tabular}{|l|l|l|l|}
\hline
\hspace*{.5cm} Mode \hspace*{.5cm}
& Data~\cite{Cheng:2006dm}
& CLF~\cite{Cheng:2006dm}
& Our results \\
\hline
$D_{s0}^{\ast -} D^{\ast 0}/D_{s0}^{\ast -} D^0$
& $0.91 \pm 0.73$ & \hspace*{.5cm} 0.49 & \hspace*{.5cm} 0.48 \\
\hline
$D_{s0}^{\ast -} D^{\ast +}/D_{s0}^{\ast -} D^+$
& $0.59 \pm 0.26$ & \hspace*{.5cm} 0.49 & \hspace*{.5cm} 0.48 \\
\hline
$D_{s1}^- D^{\ast 0}/D_{s1}^- D^0$
& $3.4 \pm 2.4$ & \hspace*{.5cm} 3.6 & \hspace*{.5cm} 2.9 \\
\hline
$D_{s1}^- D^{\ast +}/D_{s1}^- D^+$
& $2.6 \pm 1.5$ & \hspace*{.5cm} 3.6 & \hspace*{.5cm} 2.9 \\
\hline
\end{tabular}
\end{center}

\newpage

\begin{figure}

\vspace*{2cm}

\begin{center}
\epsfig{file=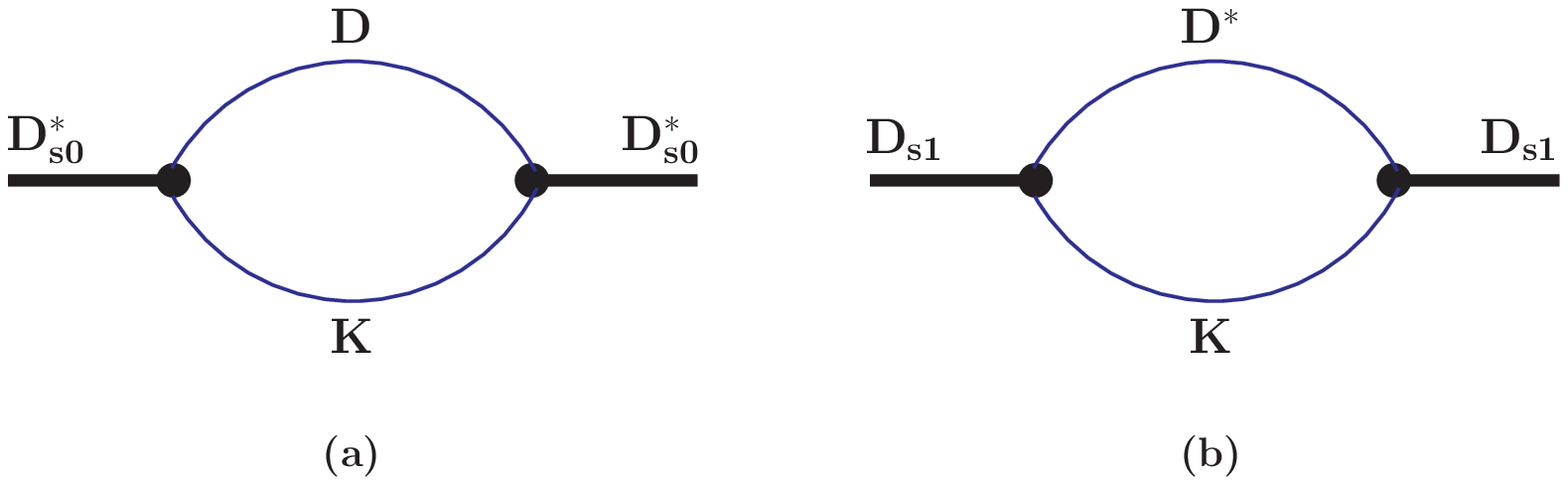, scale=.65}
\end{center}
\vspace*{.5cm}
\caption{Mass operators of
$D_{s0}^{\ast}(2317)$ and $D_{s1}(2460)$ mesons.}

\vspace*{3cm}

\begin{center}
\epsfig{file=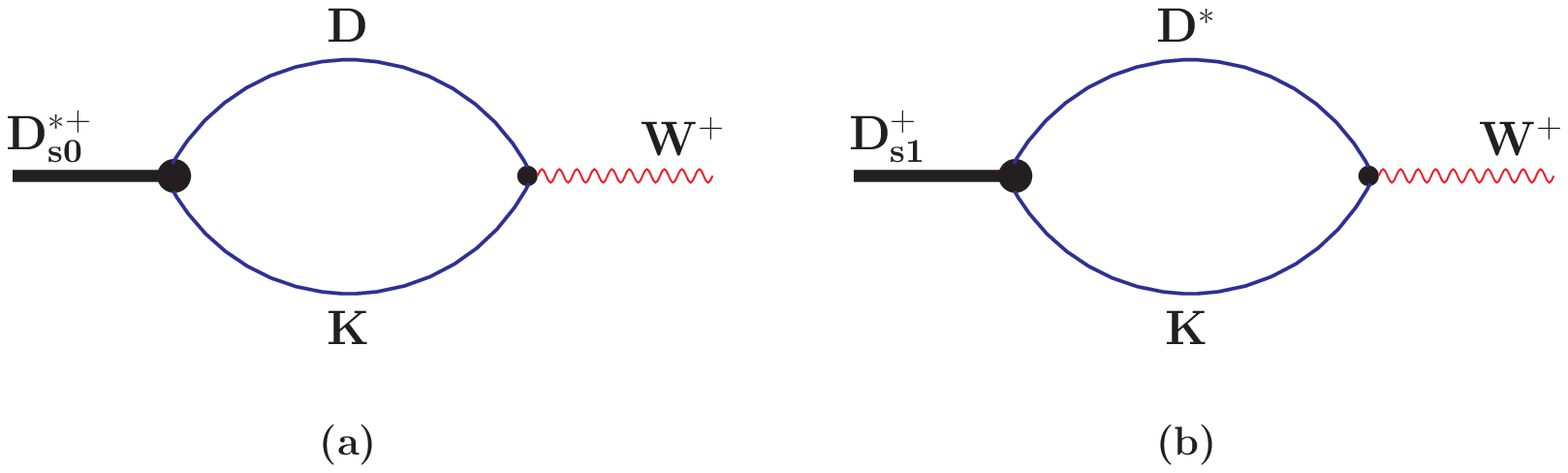, scale=.65}
\end{center}
\vspace*{.5cm}
\caption{Diagrams related to the leptonic decay constants 
of $D_{s0}^{\ast}(2317)$ and $D_{s1}(2460)$ mesons.}
\end{figure}

\begin{thebibliography}{99}
\bibitem{Rosner:2006vc}
  J.~L.~Rosner,
  Phys.\ Rev.\ D {\bf 74}, 076006 (2006)
  [arXiv:hep-ph/0608102].
\bibitem{Colangelo:2004vu}
  P.~Colangelo, F.~De Fazio and R.~Ferrandes,
  Mod.\ Phys.\ Lett.\  A {\bf 19}, 2083 (2004)
  [arXiv:hep-ph/0407137].
\bibitem{Aubert:2003fg}
  B.~Aubert {\it et al.}  [BABAR Collaboration],
  Phys.\ Rev.\ Lett.\  {\bf 90}, 242001 (2003)
  [arXiv:hep-ex/0304021].
\bibitem{Besson:2003cp}
  D.~Besson {\it et al.}  [CLEO Collaboration],
  Phys.\ Rev.\ D {\bf 68}, 032002 (2003)
  [arXiv:hep-ex/0305100].
\bibitem{Abe:2003jk}
  K.~Abe {\it et al.},
  Phys.\ Rev.\ Lett.\  {\bf 92} (2004) 012002
  [arXiv:hep-ex/0307052].
\bibitem{Krokovny:2003zq}
  P.~Krokovny {\it et al.}  [Belle Collaboration],
  Phys.\ Rev.\ Lett.\  {\bf 91}, 262002 (2003)
  [arXiv:hep-ex/0308019].
\bibitem{Aubert:2004pw}
  B.~Aubert {\it et al.}  [BABAR Collaboration],
  Phys.\ Rev.\ Lett.\  {\bf 93}, 181801 (2004)
  [arXiv:hep-ex/0408041].
\bibitem{Drutskoy:2004yv}
  A.~Drutskoy,
  J.\ Phys.\ Conf.\ Ser.\  {\bf 9}, 115 (2005)
  [arXiv:hep-ex/0412070].
\bibitem{Aubert:2006nm}
  B.~Aubert {\it et al.}  [BABAR Collaboration],
  Phys.\ Rev.\  D {\bf 74}, 031103 (2006)
  [arXiv:hep-ex/0605036].
\bibitem{Yao:2006px}
  W.~M.~Yao {\it et al.}  [Particle Data Group],
  J.\ Phys.\ G {\bf 33}, 1 (2006).
\bibitem{LeYaouanc:2001ma}
  A.~Le Yaouanc, L.~Oliver, O.~Pene, J.~C.~Raynal and V.~Morenas,
  Phys.\ Lett.\  B {\bf 520}, 59 (2001)
  [arXiv:hep-ph/0107047].
\bibitem{Cheng:2003kg}
  H.~Y.~Cheng and W.~S.~Hou,
  Phys.\ Lett.\  B {\bf 566}, 193 (2003)
  [arXiv:hep-ph/0305038].
\bibitem{Cheng:2003id}
  H.~Y.~Cheng,
  Phys.\ Rev.\  D {\bf 68}, 094005 (2003)
  [arXiv:hep-ph/0307168].
\bibitem{Hsieh:2003xj}
  R.~C.~Hsieh, C.~H.~Chen and C.~Q.~Geng,
  Mod.\ Phys.\ Lett.\  A {\bf 19}, 597 (2004)
  [arXiv:hep-ph/0312232].
\bibitem{Datta:2003re}
  A.~Datta and P.~J.~O'Donnell,
  Phys.\ Lett.\  B {\bf 572}, 164 (2003)
  [arXiv:hep-ph/0307106].
\bibitem{Datta:2004jx}
  A.~Datta,
  Int.\ J.\ Mod.\ Phys.\  A {\bf 19}, 5501 (2004)
  [arXiv:hep-ph/0407330].
\bibitem{Cheng:2003sm}
  H.~Y.~Cheng, C.~K.~Chua and C.~W.~Hwang,
  Phys.\ Rev.\  D {\bf 69}, 074025 (2004)
  [arXiv:hep-ph/0310359].
\bibitem{Hwang:2004kg}
  D.~S.~Hwang and D.~W.~Kim,
  Phys.\ Lett.\  B {\bf 606}, 116 (2005)
  [arXiv:hep-ph/0410301].
\bibitem{Thomas:2005bu}
  C.~E.~Thomas,
  Phys.\ Rev.\  D {\bf 73}, 054016 (2006)
  [arXiv:hep-ph/0511169].
\bibitem{Cheng:2006dm}
  H.~Y.~Cheng and C.~K.~Chua,
  Phys.\ Rev.\  D {\bf 74}, 034020 (2006)
  [arXiv:hep-ph/0605073].
\bibitem{Colangelo:1991ug}
  P.~Colangelo, G.~Nardulli, A.~A.~Ovchinnikov and N.~Paver,
  Phys.\ Lett.\  B {\bf 269}, 201 (1991).
\bibitem{Veseli:1996yg}
  S.~Veseli and I.~Dunietz,
  Phys.\ Rev.\  D {\bf 54}, 6803 (1996)
  [arXiv:hep-ph/9607293].
\bibitem{Colangelo:2005hv}
  P.~Colangelo, F.~De Fazio and A.~Ozpineci,
  Phys.\ Rev.\  D {\bf 72}, 074004 (2005)
  [arXiv:hep-ph/0505195].
\bibitem{Jugeau:2005yr}
  F.~Jugeau, A.~Le Yaouanc, L.~Oliver and J.~C.~Raynal,
  Phys.\ Rev.\  D {\bf 72}, 094010 (2005)
  [arXiv:hep-ph/0504206].
\bibitem{Herdoiza:2006qv}
  G.~Herdoiza, C.~McNeile and C.~Michael  [UKQCD Collaboration],
  Phys.\ Rev.\  D {\bf 74}, 014510 (2006)
  [arXiv:hep-lat/0604001].
\bibitem{Faessler:2007gv}
  A.~Faessler, T.~Gutsche, V.~E.~Lyubovitskij and Y.~L.~Ma,
  arXiv:0705.0254 [hep-ph].
\bibitem{Ds1}
A.~Faessler, Th.~Gutsche, V.~E.~Lyubovitskij, Y.~L.~Ma,
in preparation.
\bibitem{Caprini:1997mu}
  I.~Caprini, L.~Lellouch and M.~Neubert,
  Nucl.\ Phys.\  B {\bf 530}, 153 (1998)
  [arXiv:hep-ph/9712417].
\bibitem{Buchalla:1995vs}
  G.~Buchalla, A.~J.~Buras and M.~E.~Lautenbacher,
  Rev.\ Mod.\ Phys.\  {\bf 68}, 1125 (1996)
  [arXiv:hep-ph/9512380].
\bibitem{Cheng:1998kd}
  H.~Y.~Cheng and K.~C.~Yang,
  Phys.\ Rev.\  D {\bf 59}, 092004 (1999)
  [arXiv:hep-ph/9811249].
\bibitem{Luo:2001mc}
  Z.~Luo and J.~L.~Rosner,
  Phys.\ Rev.\  D {\bf 64}, 094001 (2001)
  [arXiv:hep-ph/0101089].
\bibitem{Weinberg:1962hj}
  S.~Weinberg,
  Phys.\ Rev.\  {\bf 130}, 776 (1963);
  A.~Salam,
  Nuovo Cim.\  {\bf 25}, 224 (1962);
K.~Hayashi, M.~Hirayama, T.~Muta, N.~Seto and T.~Shirafuji,
Fortsch.\ Phys.\ {\bf 15}, 625 (1967).
\bibitem{Efimov:1993ei}
G.~V.~Efimov and M.~A.~Ivanov,
{\it The Quark Confinement Model of Hadrons},
(IOP Publishing, Bristol $\&$ Philadelphia, 1993).
\bibitem{Efimov:1987sa}
  G.~V.~Efimov, M.~A.~Ivanov and V.~E.~Lyubovitskij,
  Few Body Syst.\  {\bf 6}, 17 (1989)
  [Acta Phys.\ Austriaca {\bf 6}, 17 (1989)];
  I.~V.~Anikin, M.~A.~Ivanov, N.~B.~Kulimanova and V.~E.~Lyubovitskij,
  Z.\ Phys.\ C {\bf 65}, 681 (1995);
M.~A.~Ivanov, M.~P.~Locher and V.~E.~Lyubovitskij,
Few Body Syst.\  {\bf 21}, 131 (1996);
M.~A.~Ivanov, V.~E.~Lyubovitskij, J.~G.~K\"orner and P.~Kroll,
Phys.\ Rev.\ D {\bf 56}, 348 (1997)
[arXiv:hep-ph/9612463];
  M.~A.~Ivanov and V.~E.~Lyubovitskij,
  Phys.\ Lett.\  B {\bf 408}, 435 (1997)
  [arXiv:hep-ph/9705423]; 
  M.~A.~Ivanov, J.~G.~K\"orner and V.~E.~Lyubovitskij,
  Phys.\ Lett.\ B {\bf 448}, 143 (1999)
  [arXiv:hep-ph/9811370];
  M.~A.~Ivanov, J.~G.~K\"orner, V.~E.~Lyubovitskij and A.~G.~Rusetsky,
  Phys.\ Rev.\ D {\bf 60}, 094002 (1999)
  [arXiv:hep-ph/9904421];
  M.~A.~Ivanov and P.~Santorelli,
  Phys.\ Lett.\ B {\bf 456}, 248 (1999)
  [arXiv:hep-ph/9903446];
  A.~Faessler, T.~Gutsche, M.~A.~Ivanov, J.~G.~K\"orner
  and V.~E.~Lyubovitskij,
  Phys.\ Lett.\ B {\bf 518}, 55 (2001)
  [arXiv:hep-ph/0107205];
  A.~Faessler, T.~Gutsche, M.~A.~Ivanov, V.~E.~Lyubovitskij and P.~Wang,
  Phys.\ Rev.\  D {\bf 68}, 014011 (2003)
  [arXiv:hep-ph/0304031].
  M.~A.~Ivanov, J.~G.~Korner and P.~Santorelli,
  Phys.\ Rev.\ D {\bf 73}, 054024 (2006)
  [arXiv:hep-ph/0602050];
A.~Faessler, T.~Gutsche, M.~A.~Ivanov, J.~G.~Korner,
V.~E.~Lyubovitskij, D.~Nicmorus and K.~Pumsa-ard,
Phys.\ Rev.\ D {\bf 73}, 094013 (2006)
[arXiv:hep-ph/0602193];
  A.~Faessler, T.~Gutsche, B.~R.~Holstein, V.~E.~Lyubovitskij,
  D.~Nicmorus and K.~Pumsa-ard,
  Phys.\ Rev.\ D {\bf 74}, 074010 (2006)
  [arXiv:hep-ph/0608015].
\bibitem{Burdanov:1996uw}
  Y.~V.~Burdanov, G.~V.~Efimov, S.~N.~Nedelko and S.~A.~Solunin,
  Phys.\ Rev.\ D {\bf 54}, 4483 (1996)
  [arXiv:hep-ph/9601344]; 
  I.~V.~Anikin, A.~E.~Dorokhov and L.~Tomio,
  Phys.\ Part.\ Nucl.\  {\bf 31}, 509 (2000)
  [Fiz.\ Elem.\ Chast.\ Atom.\ Yadra {\bf 31}, 1023 (2000)].
\bibitem{Burdanov:2000rw}
  J.~V.~Burdanov and G.~V.~Efimov,
  Phys.\ Rev.\ D {\bf 64}, 014001 (2001)
  [arXiv:hep-ph/0009027].
\bibitem{Baru:2003qq}
  V.~Baru, J.~Haidenbauer, C.~Hanhart, Yu.~Kalashnikova and A.~E.~Kudryavtsev,
  Phys.\ Lett.\  B {\bf 586}, 53 (2004)
  [arXiv:hep-ph/0308129]; 
  C.~Hanhart, Yu.~S.~Kalashnikova, A.~E.~Kudryavtsev and A.~V.~Nefediev,
  Phys.\ Rev.\  D {\bf 75}, 074015 (2007)
  [arXiv:hep-ph/0701214].
\bibitem{Wirbel:1985ji}
  M.~Wirbel, B.~Stech and M.~Bauer,
  Z.\ Phys.\  C {\bf 29}, 637 (1985).
\bibitem{Flynn:1997ca}
  J.~M.~Flynn and C.~T.~Sachrajda,
  Adv.\ Ser.\ Direct.\ High Energy Phys.\  {\bf 15}, 402 (1998)
  [arXiv:hep-lat/9710057].
\bibitem{Ball:2006yd}
  P.~Ball,
  Phys.\ Lett.\  B {\bf 641}, 50 (2006)
  [arXiv:hep-ph/0608116].
\bibitem{Zweber:2007dy}
  P.~Zweber,
  arXiv:hep-ex/0701018.
\bibitem{Neubert:1993mb}
  M.~Neubert,
  Phys.\ Rept.\  {\bf 245}, 259 (1994)
  [arXiv:hep-ph/9306320].
\bibitem{Falk:1992wt}
  A.~F.~Falk and M.~Neubert,
  Phys.\ Rev.\  D {\bf 47}, 2965 (1993)
  [arXiv:hep-ph/9209268];
  Phys.\ Rev.\  D {\bf 47}, 2982 (1993)
  [arXiv:hep-ph/9209269].
\bibitem{HFAG:2006}
 Heavy Flavor Averaging Group (HFAG), http://www.slac.stanford.edu/xorg/hfag/

\end{thebibliography}
\end{document}